\documentclass[journal,comsoc]{IEEEtran}
\usepackage[T1]{fontenc}

\usepackage{cite}

\usepackage{subfigure}
   \usepackage[pdftex]{graphicx}
\usepackage{amsmath}
\usepackage{amssymb}
\usepackage{amsthm}
\usepackage{amsmath}
\usepackage{extarrows}
\usepackage{color}
\usepackage[cmintegrals]{newtxmath}

\usepackage{hyperref}
\begin{document}
%
\title{On the Performance of NOMA-Based Cooperative Relaying Systems over Rician Fading Channels}
%
%
%

\author{Ruicheng~Jiao,
        Linglong~Dai,
        Jiayi Zhang,
         Richard MacKenzie, and Mo Hao
\thanks{Copyright (c) 2015 IEEE. Personal use of this material is permitted.
However, permission to use this material for any other purposes must be
obtained from the IEEE by sending a request to pubs-permissions@ieee.org.}
\thanks{This work was supported in part by the International Science \& Technology Cooperation Program of China (Grant No. 2015DFG12760),  
the National Natural Science Foundation of China (Grant No. 61571270), the Royal Academy of
Engineering under the UK-China Industry Academia Partnership Programme
Scheme (Grant No. UK-CIAPP $\backslash$49), and the British
Telecom and Tsinghua SEM Advanced ICT LAB.}
\thanks{R. Jiao and L. Dai are with the Tsinghua National Laboratory for Information Science and Technology (TNList), Department
of Electronic Engineering, Beijing 100084, China (e-mails: jiaors16@mails.tsinghua.edu.cn, daill@tsinghua.edu.cn).}
\thanks{J. Zhang is with the School of Electronics and Information Engineering, Beijing Jiaotong University, Beijing 100044, China (e-mail: jiayizhang@bjtu.edu.cn).}
\thanks{R. MacKenzie is with the BT TSO, Adastral Park, Ipswich, U.K. (e-mail: richard.mackenzie@bt.com).}
\thanks{M. Hao is with the Tsinghua SEM Advanced ICT LAB, Tsinghua University,
Beijing 100084, China (e-mail: haom@sem.tsinghua.edu.cn).}
}

\maketitle

\begin{abstract}
Non-orthogonal multiple access (NOMA) is a promising technique for the fifth generation (5G) wireless communications. As users with good channel conditions can serve as relays to enhance the system performance by using successive interference cancellation (SIC), the integration of NOMA and cooperative relaying has recently attracted increasing interests. In this paper, a NOMA-based cooperative relaying system is studied, and an analytical framework is developed to evaluate its performance. Specifically, the performance of NOMA over Rician fading channels is studied, and the exact expression of the average achievable rate is derived. Moreover, we also propose an approximation method to calculate the achievable rate by using the Gauss-Chebyshev Integration. Numerical results confirm that our derived analytical results match well with the Monte Carlo simulations.
\end{abstract}


%
\IEEEpeerreviewmaketitle

\vspace*{-3mm}
\begin{figure*}[!t]
\centering
\subfigure[]
{
  \includegraphics[height=1.5in]{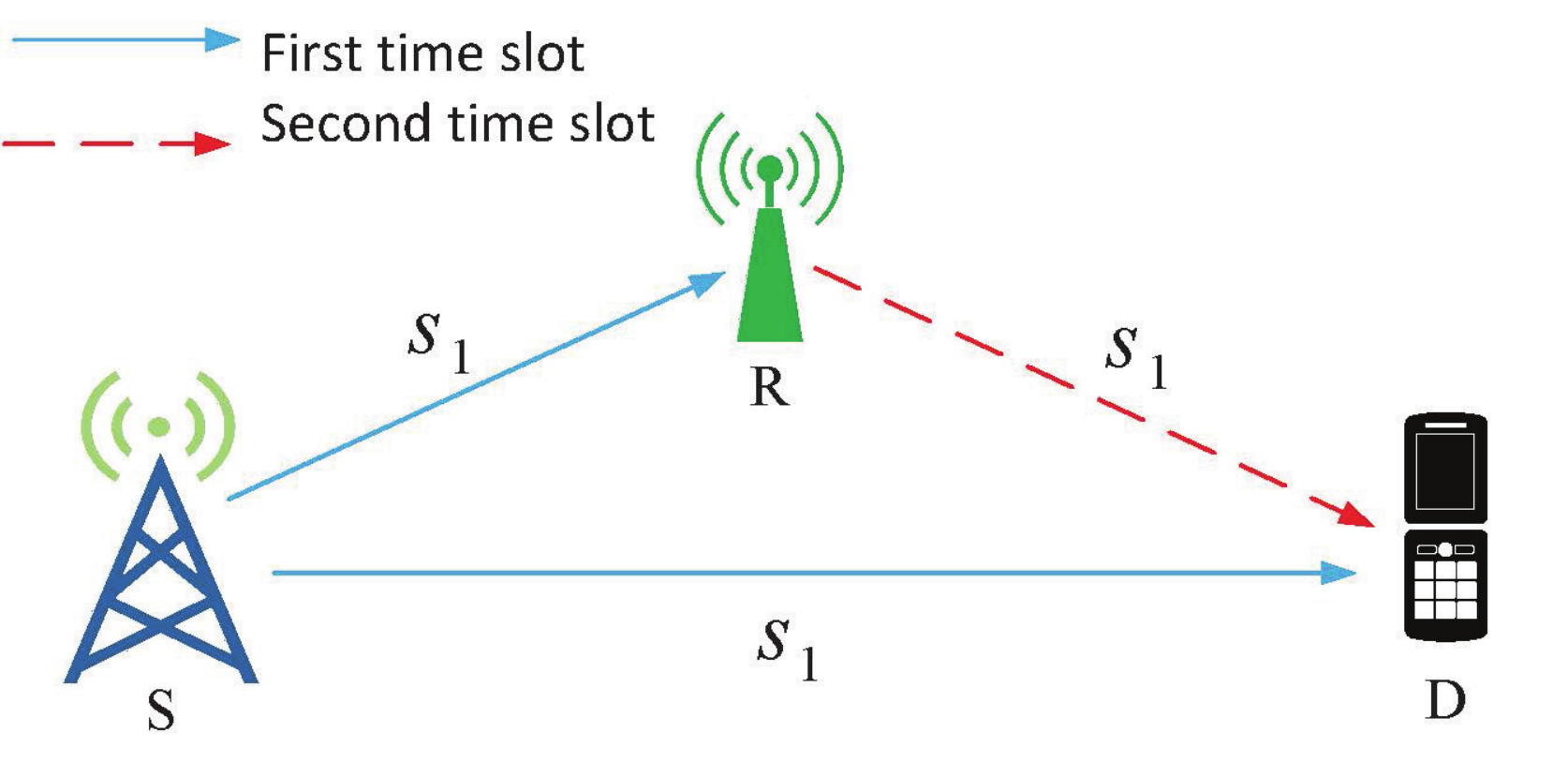}

  \label{system_model_1}
   \vspace*{-2mm}
}
\subfigure[]
{
  
  \includegraphics[height=1.5in]{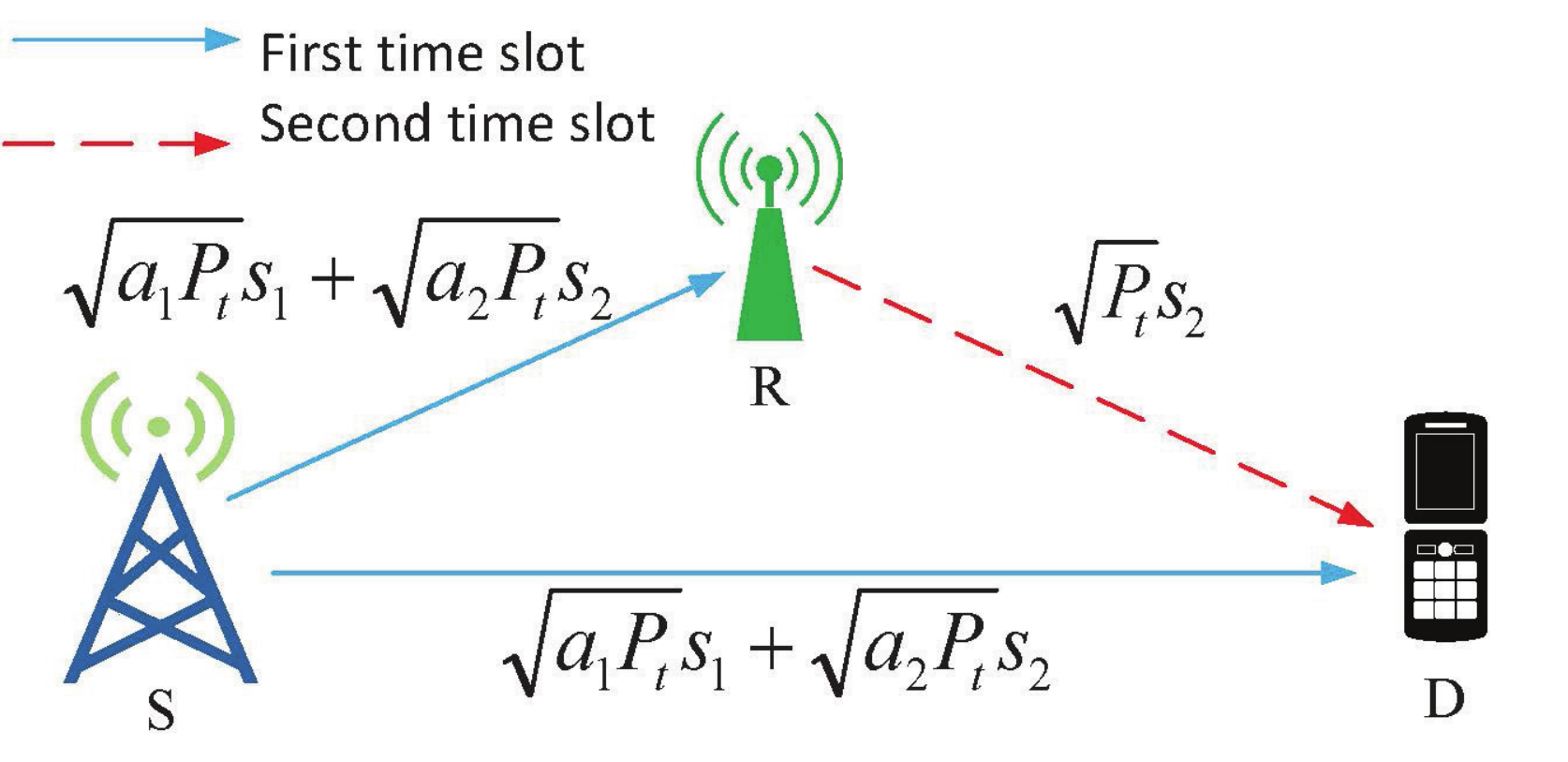}
  \label{system_model_2}
  \vspace*{-2mm}
}
\caption{System models of two cooperative relaying systems: (a) Traditional cooperative relaying systems; (b) NOMA-based cooperative relaying systems.}
\label{system_model}
\vspace*{-5mm} 
\end{figure*}

\section{Introduction}

%
%
%
%

\IEEEPARstart{I}{t} is highly expected that future 5G networks should achieve a 10-fold increase in connection density, i.e., $10^6$ connections per square kilometers \cite{boccardi2014five}. Non-orthogonal multiple access (NOMA) has been proposed as a promising candidate to realize such an aggressive 5G goal \cite{dai2015non, fang2016efficient, sun2016allocation, saito2013system}. NOMA is foundamentally different from conventional orthogonal multiple access (OMA) schemes such as FDMA, TDMA, OFDMA, etc., since it allows multiple users to simultaneously transmit signals using the same time/frequency radio resources but different power levels \cite{fang2016efficient, sun2016allocation, saito2013system}. The key advantage of NOMA is to explore the extra power domain to further increase the number of supportable users. Specifically, users are identified by their channel conditions, those with good channel conditions are called strong users and others are called weak users. For the sake of fairness, less power are allocated to strong users at the transmitter side. In this way, the transmitter sends the superposition of signals with different power levels and the receiver applies successive interference cancellation (SIC) to strong users to realize multi-user detection\cite{saito2013system, liu2016}. Such non-orthogonal resource allocation enables NOMA to accomodate more users and makes it promising to address the 5G requirement of massive connectivity, with the cost of controllable increase of complexity in receiver design due to SIC\cite{saito2013system}.

In NOMA systems, the use of SIC implies that strong users have prior information about the messages of other users, so essentially they are able to serve as cooperative relays. Moreover, cooperative relaying is able to significantly enhance the system performance of cellular networks \cite{laneman2004cooperative}. Thus, combining cooperative relaying and NOMA is promising to improve the throughput of future 5G wireless networks, and has attracted increasing interests recently\cite{liu2015cooperative}. Specifically, a cooperative NOMA transmission scheme was proposed in \cite{ding2015cooperative}, where strong users decode the signals that are intended to others and serve as relays to improve the performance of weak users. Another NOMA-based cooperative scheme was proposed in \cite{kim2015capacity}, where the performance of a NOMA-based decode-and-forward relaying system under Rayleigh fading channel was studied.
However, most of existing NOMA schemes only consider the Rayleigh fading channel, which is suitable for rich scattering scenarios without line of sight (LOS), while little attention has been drawn to the more general Rician fading channel, which takes both LOS and non LOS (NLOS) into consideration. In some typical 5G application scenarios, such as massive machine-type communications (mMTC) and Internet of things (IoT),  ``users'' may be low-cost sensors deployed in a small area, where both LOS and NLOS exist, which can be better modeled
by the Rician fading channel.

In this paper, we investigate the performance of the NOMA-based cooperative relaying transmission scheme in \cite{kim2015capacity} under Rician fading channels\footnote{Simulation codes are provided to reproduce the results presented in this paper: \url{http://oa.ee.tsinghua.edu.cn/dailinglong/publications/publications.html}.}. Evaluating system performance under Rician fading channel is rather challenging as the probability density function of Rician distribution variables consists of Bessel function, which makes it difficult to calculate the average achievable rate through integration. In order to derive the exact expression of the achievable rate, we propose an analytical method using Taylor expansion of Bessel function and incomplete Gamma function. However, the complexity of the incomplete Gamma function makes it still difficult to get the exact values, so we further propose an approximation method using Gauss-Chebvshev Integration to simplify the calculation. Finally, simulations confirm that our analytical results match well with the Monte Carlo results.

The rest of the paper is organized as follows. The system model of the NOMA-based cooperative relaying system is introduced in Section II. In Section III, we provide a detailed
analysis for the achievable rate of the system and acquire both accurate and approximated results. Section IV provides numerical results to validate the theoretical analysis and section V concludes the paper.

\vspace*{-4mm}

\section{System Model}

As illustrated in Fig. \ref{system_model} (a) and Fig. \ref{system_model} (b), we consider a simple cooperative relaying system (CRS) consisting of a source (S), a decode-and-forward relay (R) which works in half-duplex mode, and a destination (D).
We assume that all links between them (i.e., S-to-D, S-to-R, and R-to-D) are available. The independent Rician fading channel coefficients of S-to-D, S-to-R, and R-to-D links are denoted as $h_{\rm{SD}}$, $h_{\rm{SR}}$, and $h_{\rm{RD}}$, with the average powers of $\Omega_{\rm{SD}}$, $\Omega_{\rm{SR}}$, and $\Omega_{\rm{RD}}$, respectively. It is also assumed that $\Omega_{\rm{SD}} <\Omega_{\rm{SR}}$, since in general the path loss of the S-to-D link is usually worse than that of the S-to-R link \cite{kim2015capacity}. 

In the traditional CRS presented in Fig. \ref{system_model} (a), the source transmits $s_1$ to the relay and destination in the first time slot. Then in the second time slot, the relay transmits $s_1$ to the destination. In this way, the destination only receives one signal in two time slots. 

In the NOMA-based CRS showed in Fig. \ref{system_model} (b), the destination is able to receive two different signals in two time slots, so it outperforms the traditional CRS in terms of throughput. Specifically, in the first time slot, the source transmits the superposition of two different data symbols $s_1$ and $s_2$ to the relay and the destination as follows:
\begin{align}
t=\sqrt{a_1 P_t}s_1+\sqrt{a_2 P_t}s_2,
\end{align}

where $s_i$ denotes the \emph{i}-th data symbol with normalized power $E[|s_i|^2]=1$, $P_t$ is the total transmit power, and $a_i$ is the power allocation coefficient. It is noted that $a_1+a_2=1$, and $a_1>a_2$ due to  $\Omega_{\rm{SD}}^2 <\Omega_{\rm{SR}}^2$ \cite{saito2013system}. Thus, the received signals $r_{\rm{SR}}$ and $r_{\rm{SD}}$ at the relay and the destination in the first time slot  are respectively expressed as 
\begin{align}
\label{receive_1} r_{\rm{SR}}=h_{\rm{SR}}(\sqrt{a_1 P_t}s_1+\sqrt{a_2 P_t}s_2)+n_{\rm{SR}}, \\
\label{receive_2} r_{\rm{SD}}=h_{\rm{SD}}(\sqrt{a_1 P_t}s_1+\sqrt{a_2 P_t}s_2)+n_{\rm{SD}},
\end{align}

where $n_{\rm{SR}}$ and $n_{\rm{SD}}$ denote the additive white Gaussian noise (AWGN) with zero mean and variance $\sigma^2$. The destination only decodes symbol $s_1$ by treating symbol $s_2$ as noise, while the relay acquires symbol $s_2$ from (1) using SIC. Thus, the received signal-to-interference plus noise ratios (SINRs) for symbols $s_1$ and $s_2$ at the relay can be respectively obtained as
\begin{align}
\label{SNR_1} \gamma_{\rm{SR}}^1&=\frac{|h_{\rm{SR}}|^2 a_1 P_t}{|h_{\rm{SR}}|^2 a_2 P_t+\sigma^2}, \\
\label{SNR_2} \gamma_{\rm{SR}}^2&=\frac{|h_{\rm{SR}}|^2 a_2 P_t}{\sigma^2},
\end{align} 
and the received SINR for symbol $s_1$ at the destination is obtained as
\begin{align}
\label{SNR_3} \gamma_{\rm{SD}}=\frac{|h_{\rm{SD}}|^2 a_1 P_t}{|h_{\rm{SD}}|^2 a_2 P_t+\sigma^2}.
\end{align}
In the second time slot, only the relay transmits the decoded symbol $s_2$ with full power $P_t$ to the destination. Assuming that the relay can perfectly decode symbol $s_2$ in the first time slot\cite{saito2013system}, the received signal at the destination in the second time slot can be expressed as 
\begin{align}
\label{receive_3} r_{\rm{RD}}=h_{\rm{RD}}\sqrt{P_t} s_2+n_{RD},
\end{align}
where $n_{\rm{RD}}$ is the AWGN with zero mean and variance $\sigma^2$, and the received SINR for symbol $s_2$ in (\ref{receive_3}) can be obtained as
\vspace*{-6mm}

\begin{align}
\label{SNR-4} \gamma_{\rm{RD}}=\frac{|h_{\rm{RD}}|^2 P_t}{\sigma^2}.
\end{align}
 
As the expressions for received signals and SINRs are already acquired, we will calculate both the exact and approximated achievable rates in the NOMA-based CRS in the next section.

\vspace*{-2mm}
\section{Achievable Rate Analysis and Approximation}

In this section, we first derive the exact expression of the average achievable rate of the NOMA-based CRS over Rician fading channel. As the exact value of achievable rates are difficult to calculate, we further propose an approximation method using Gauss-Chebyshev Integration to simplify the numerical calculation.

\vspace*{-3mm}
\subsection{Achievable Rate Analysis}

In this subsection, we analyze the average achievable rate of $s_1$ and $s_2$. Let $\lambda_{\rm{SD}} \triangleq |h_{\rm{SD}}|^2$, $\lambda_{\rm{SR}} \triangleq |h_{\rm{SR}}|^2$, 
$\lambda_{\rm{RD}} \triangleq |h_{\rm{RD}}|^2$, and $\rho \triangleq P_t/\sigma^2$, where $\rho$ represents the transmit SNR. 
As both the relay and the destination must successfully decode $s_1$ and $s_2$, the rates of these two signals should be lower than the rates of both links calculated by Shanon formula, so the achievable rate is the minimum of the rates of two different links.
According to \cite{kim2015capacity}, we can obtain the achievable rates $C_{s_1}$ and $C_{s_2}$ of signals $s_1$ and $s_2$ respectively as
\begin{align}
\label{ave_1}C_{s_1}&=\frac{1}{2} \min\left\{\log_2(1+\gamma_{\rm{SD}}),\log_2(1+\gamma_{\rm{SR}}^1)\right\}  \nonumber \\ 
       &=\frac{1}{2}\log_2\Big(1+\min\{\lambda_{\rm{SD}}, \lambda_{\rm{SR}}\}\rho\Big)-             \nonumber \\  
       &\quad \,\,  \frac{1}{2}\log_2\Big(1+\min\{\lambda_{\rm{SD}}, \lambda_{\rm{SR}}\}\rho a_2\Big),
\end{align}

\begin{align}
\label{ave-2}C_{s_2}&=\frac{1}{2} \min\Big\{\log_2(1+\gamma_{\rm{SR}}^2),\log_2(1+\gamma_{\rm{RD}})\Big\}  \nonumber \\ 
       &=\frac{1}{2}\log_2\Big(1+\min\{a_2\lambda_{\rm{SR}}, \lambda_{\rm{RD}}\}\rho\Big).                     
\end{align}
Let $z_1\triangleq \min\{\lambda_{\rm{SR}}, \lambda_{\rm{SD}}\}$, $z_2 \triangleq \min\{a_2\lambda_{\rm{SR}}, \lambda_{\rm{RD}}\}$. According to \cite{bhatnagar2013capacity}, we can get the cumulative distribution function (CDF) of $z_1$ as
\begin{align}
\label{CDF_1}F(z_1)&=1-A_xA_y\sum_{k=0}^\infty\sum_{n=0}^\infty\widetilde{B}_x(n)\widetilde{B}_y(k)\Gamma(n+1,a_xz_1)   \nonumber  \\
&~~~\times \Gamma(k+1,a_yz_1)     \nonumber  \\  
&\overset{(a)}{=}1-A_xA_y \sum_{k=0}^\infty\sum_{n=0}^\infty \widetilde{B}_x(n) \widetilde{B}_y(k)n!k!e^{-(a_x+a_y)z_1}        \nonumber  \\
&~~~~\times\sum_{i=0}^n \sum_{j=0}^k\frac{a_x^i a_y^j}{i!j!}z^{i+j}_1,
\end{align}

where $B_x(n)=(K_x^n(1+K_x)^n)/(\Omega_x^n(n!)^2)$, $B_y(k)=(K_y^k(1+K_y)^k)/(\Omega_y^k(k!)^2)$, $a_x=(1+K_x)/\Omega_x$, $a_y=(1+K_y)/\Omega_y$, $A_x=a_xe^{-K_x}$, $A_y=a_ye^{-K_y}$, $\widetilde{B}_x(n)=B_x(n)/a_x^{n+1}$, $\widetilde{B}_y(k)=B_y(k)/a_y^{k+1}$. The subscript $x$ denotes the S-to-D link, $y$ denotes the S-to-R link, $w$ denotes the R-to-D link, and $K$ is the Rician factor. Note that the expansion form of incomplete Gamma function is used for the second equality (a) of (\ref{CDF_1}). 

Then, we prove the convergence of the infinite summation in (\ref{CDF_1}) as follows.
\begin{IEEEproof}
Let $P_x=(K_x(1+K_x))/\Omega_x$, $Q_y=(K_y(1+K_y))/\Omega_y$, we have
\begin{align}
\frac{\Gamma(n+1,a_xz_1)}{n!}<\frac{\Gamma(n+1)}{n!}<1,  \\
\frac{\Gamma(k+1,a_yz_1)}{n!}<\frac{\Gamma(n+1)}{n!}<1, 
\end{align}
then
\begin{align}
&A_xA_y \sum_{k=0}^\infty\sum_{n=0}^\infty \widetilde{B}_x(n) \widetilde{B}_y(k) \Gamma(n+1,a_xz_1)\Gamma(k+1,a_yz_1)  \nonumber  \\
&=A_xA_y \sum_{k=0}^\infty\sum_{n=0}^\infty \frac{P_x^n}{n!}\frac{Q_y^k}{k!}  \frac{\Gamma(n+1,a_xz_1)}{n!}\frac{\Gamma(k+1,a_yz_1)}{k!}    \nonumber  \\
&<A_xA_y\sum_{k=0}^\infty\sum_{n=0}^\infty\frac{P_x^n}{n!}\frac{Q_y^k}{k!}   \nonumber \\
&=A_xA_ye^{P_x+Q_y}.      \nonumber
\end{align}
The final value will not change as $k$ or $n$ increases, so the infinite summation in (\ref{CDF_1}) is convergent. 
\end{IEEEproof}

 Similarly, we can obtain the CDF of $z_2$ as follows:
 \vspace*{-2mm}
\begin{align}
\label{CDF_2} G(z_2)=&1-A_wA_y \sum_{n=0}^\infty\sum_{k=0}^\infty \widetilde{B}_w(n) \widetilde{B}_y(k) \Gamma(n+1,a_wz_2)  \nonumber \\
&\times \Gamma(k+1,\frac{a_y}{a_2}z_2)    \nonumber \\
=&1-A_wA_y \sum_{k=0}^\infty\sum_{n=0}^\infty \widetilde{B}_w(n) \widetilde{B}_y(k)n!k!e^{-(a_w+\frac{a_y}{a_2})z_2}        \nonumber  \\
&\times\sum_{i=0}^n \sum_{j=0}^k\frac{a_w^i (\frac{a_y}{a_2})^j}{i!j!}z^{i+j}_2,
\end{align}
where the parameters in (\ref{CDF_2}) are similarly defined as those in (\ref{CDF_1}).

After the CDF of $z_1 \triangleq \min\{\lambda_{\rm{SR}},\lambda_{\rm{SD}}\}$ has been obtained as (\ref{CDF_1}), we can substitute it into (\ref{ave_1}), and then the average achievable rate $C_{s_1}$ of the signal $s_1$ as shown in (\ref{ave_1}) can be expressed as 
\begin{align}
\label{R_1} C_{s_1}&=\frac{1}{2}\int_{0}^{\infty}{ [ \log_2(1+z_1\rho)-\log_2(1+z_1\rho a_2)]dF(z_1)}  \nonumber \\
&=\frac{1}{2\ln(2)}\Bigg[\rho\int_{0}^{\infty}{\frac{1-F(z_1)}{1+z_1\rho}dz_1}-\rho a_2\int_{0}^{\infty}{\frac{1-F(z_1)}{1+z_1\rho a_2}dz_1}\Bigg].
\end{align}

\vspace*{-4mm}
Let $\displaystyle D(\rho)=\rho\int_{0}^{\infty}{\frac{1-F(z_1)}{1+z_1\rho}dz_1}$, and substitute (\ref{CDF_1}) into $D(\rho)$, we have
\begin{align}
\label{D} D(\rho)&=\rho\int_{0}^{\infty}{\frac{1-F(z_1)}{1+z_1\rho}dz_1}   \nonumber   \\
&=A_xA_y \sum_{k=0}^\infty\sum_{n=0}^\infty \widetilde{B}_x(n) \widetilde{B}_y(k)n!k!\times\sum_{i=0}^n \sum_{j=0}^k\frac{a_x^i a_y^j}{i!j!}        \nonumber  \\
&~~~\int_0^\infty \frac{z^{i+j}_1e^{-(a_x+a_y)z_1}}{1+z_1\rho}d(z_1\rho)  \nonumber  \\
&\overset{(b)}{=}A_xA_y \sum_{k=0}^\infty\sum_{n=0}^\infty \widetilde{B}_x(n) \widetilde{B}_y(k)n!k!\times\sum_{i=0}^n \sum_{j=0}^k\frac{a_x^i a_y^j}{i!j!\rho^{i+j}}       \nonumber  \\
&~~~~\int_0^\infty \frac{t^{i+j}e^{-\frac{a_x+a_y}{\rho}t}}{1+t}dt,
\end{align}
where ($b$) is obtained by setting $t=z_1\rho$.

Now we have the following \textbf{Lemma 1} to calculate the integral $\displaystyle\int_0^\infty \frac{t^{i+j}e^{-\frac{a_x+a_y}{\rho}t}}{1+t}dt$ in (\ref{D}).
\newtheorem{lemma}{Lemma}
\begin{lemma}
For $m \in \mathbb{Z}^*$ and $\beta > 0$, we have 
\begin{align}
\label{gamma} \int_0^\infty\frac{t^m e^{-\beta t}}{1+t}dt=e^\beta m! \Gamma(-m,\beta),
\end{align}

\vspace*{-1mm}
where  $\displaystyle \Gamma(-m,\beta)=\int_\beta^\infty\frac{e^{-t}}{t^{m+1}}dt$ denotes the incomplete Gamma function. 
\end{lemma}
\vspace*{-1mm}

\begin{IEEEproof}
Let $x=\beta(1+t)$, we have 
\begin{align}
\label{lemma} \int_0^\infty\frac{t^m e^{-\beta t}}{1+t}dt=\frac{e^\beta}{\beta^m}\int_\beta^\infty\frac{(x-\beta)^me^{-x}}{x}dx.
\end{align}
Then we define:
\begin{align}
\label{proof_1} J_m(x)&=\frac{(x-\beta)^m}{x},           \\
\label{proof_2}I(x)&=\frac{e^\beta}{\beta^m}\int_\beta^\infty J_m(x)e^{-x}dx.
\end{align}

On the one hand, by substituting (\ref{proof_1}) into (\ref{proof_2}), we have
\begin{align}
\label{proof} I(x)&=-\frac{e^\beta}{\beta^m}\int_\beta^\infty\frac{(x-\beta)^m}{x}d(e^{-x})     \nonumber  \\
&=-\frac{e^\beta}{\beta^m}\frac{(x-\beta)^m}{x}e^{-x}\Big|_\beta^\infty+\frac{e^\beta}{\beta^m} \times  \nonumber   \\
&~~~\int_\beta^\infty\frac{(x-\beta)^{m-1}(mx-x-\beta)}{x^2}e^{-x}dx               \nonumber  \\
&=\frac{e^\beta}{\beta^m}\int_\beta^\infty J_{m-1}(x)e^{-x}dx.    
\end{align}

We can observe from (\ref{proof}) that as long as $\beta$ is a root of $J_m(x)$, the integral can be successively calculated by using (\ref{proof}) for $m$ times. On the other hand, we know that $J_m(x)=(x-\beta)^m/x=x^{m-1}+a_{m-2}x^{m-2}+...+a_0+(-1)^m\beta^m/x$, so after $m$ times of integration by part, we will have
\begin{align}
I(x)&=-\frac{e^\beta}{\beta^m}\int_\beta^\infty\frac{(x-\beta)^m}{x}d(e^{-x})    \nonumber  \\
&= -\frac{e^\beta}{\beta^m}\int_\beta^\infty \Big(x^{m-1}+...+a_0+(-1)^m\frac{\beta^m}{x}\Big)^{(m)}d(e^{-x})     \nonumber  \\
&= -\frac{e^\beta}{\beta^m}\int_\beta^\infty \Big((-1)^m\frac{\beta^m}{x}\Big)^{(m)}d(e^{-x})   \nonumber    \\
&=e^\beta m! \Gamma(-m,\beta), 
\end{align}

where $(\cdot)^{(m)}$ denotes $m$-order derivation. 
\end{IEEEproof}

Substitute (\ref{gamma}) in \textbf{Lemma 1} into (\ref{D}), we can get the final exact expression of $C_{s_1}$ as
\begin{align}
\label{c_1} C_{s_1}=\frac{1}{2\ln(2)}(D(\rho)-D(\rho a_2)),
\end{align}
where 
\begin{align}
D(\rho)&=A_xA_y \sum_{k=0}^\infty\sum_{n=0}^\infty \widetilde{B}_x(n) \widetilde{B}_y(k) n!k!       \nonumber  \\
&~~~\times\sum_{i=0}^n \sum_{j=0}^k\frac{(i+j)!}{i!j!}\frac{a_x^i a_y^j}{\rho^{i+j}}e^{\frac{a_x +a_y}{\rho}}\Gamma(-i-j,\frac{a_x +a_y}{\rho}),
\end{align}

and $D(\rho a_2)$ shares the same form as $D(\rho)$.

Similarly, we can derive the exact expression of $C_{s_2}$ as
\begin{align}
\label{c_2} C_{s_2}&=\frac{1}{2\ln(2)}A_wA_y \sum_{k=0}^\infty\sum_{n=0}^\infty \widetilde{B}_w(n) \widetilde{B}_y(k) n!k!\times\sum_{i=0}^n \sum_{j=0}^k      \nonumber  \\
~~~& \times \frac{(i+j)!}{i!j!} \frac{a_w^i (a_y/a_2)^j}{\rho^{i+j}} e^{\frac{a_w+a_y/a_2}{\rho}}\Gamma(-i-j,\frac{a_w+a_y/a_2}{\rho}).
\end{align}

Although we have derived the exact expressions of the achievable rates of $s_1$ and $s_2$ in (\ref{c_1}) and (\ref{c_2}) respectively, such expressions are very complicated, since the incomplete Gamma function is difficult to calculate. Thus, it is still difficult to get the exact values of the achievable rates, which motivates us to  propose an approximation method to solve this problem in the next subsection.

\vspace*{-4mm}
\subsection{Achievable Rate Approximation }

In this subsection, we propose an approximation method using Gauss-Chebyshev Integration \cite{book} to simplify the numerical calculation of the incomplete Gamma function $\Gamma(-m,\beta)$. 
 However, Gauss-Chebyshev Integration is used on the limited interval $[-1,1]$, while the integral intervals in incomplete Gamma functions of (\ref{c_1}) and (\ref{c_2}) are infinite intervals. Thus, we set $t=2\beta\frac{1}{x}-1$ and convert the incomplete Gamma function as
\begin{align}
\label{appro} \Gamma(-m,\beta)&=(\frac{1}{2 \beta})^m \int_{-1}^{1}\frac{1}{\sqrt{1-t^2}}(t+1)^{m-1}e^{-\frac{2 \beta}{t+1}}\sqrt{1-t^2}dt  \nonumber  \\
&=(\frac{1}{2 \beta})^m \frac{\pi}{n} \sum_{l=1}^n (\cos(\frac{2l-1}{2n}\pi)+1)^{m-1}  \nonumber  \\
&~~~\times e^{-\frac{2 \beta}{\cos(\frac{2l-1}{2n}\pi)+1}}  |\sin(\frac{2l-1}{2n}\pi)|,
\end{align}
where $n$ is the approximation order. Substituting (\ref{appro}) into the exact expression of the achievable rate (\ref{c_1}), we can finally obtain the approximation of (\ref{c_1}) as
\begin{align}
\label{c_1_appro} C_{s_1}=\frac{1}{2\ln(2)}\Big(D(\rho)-D(\rho a_2)\Big),
\end{align}
\vspace*{-2mm}
where 
\vspace*{-2mm}
\begin{align}
D(\rho)&=A_xA_y \sum_{k=0}^\infty\sum_{n=0}^\infty \widetilde{B}_x(n) \widetilde{B}_y(k) n!k!\times\sum_{i=0}^n \sum_{j=0}^k\frac{(i+j)!}{i!j!}\frac{a_x^i a_y^j}{\rho^{i+j}}       \nonumber  \\
&~~~\times e^{\frac{a_x +a_y}{\rho}}\Bigg(\frac{1}{2 \frac{a_x +a_y}{\rho}}\Bigg)^{i+j} \frac{\pi}{n} \sum_{l=1}^n (\cos(\frac{2l-1}{2n}\pi)+1)^{i+j-1}  \nonumber  \\
&~~~\times   e^{-\frac{2 \frac{a_x +a_y}{\rho}}{\cos(\frac{2l-1}{2n}\pi)+1}}  |\sin(\frac{2l-1}{2n}\pi)|,
\end{align}
and $D(\rho a_2)$ shares the same form as $D(\rho)$. Similarly, (\ref{c_2}) can be approximated as
\begin{align}
\label{c_2_appro} C_{s_2}&=\frac{1}{2\ln(2)}A_wA_y \sum_{k=0}^\infty\sum_{n=0}^\infty \widetilde{B}_w(n) \widetilde{B}_y(k) n!k! \sum_{i=0}^n\sum_{j=0}^k\frac{(i+j)!}{i!j!} \nonumber  \\
&~~~\times  \frac{a_w^i (a_y/a_2)^j}{\rho^{i+j}}  e^{\frac{a_w +a_y/a_2}{\rho}} \Bigg(\frac{1}{2 \frac{a_w +a_y/a_2}{\rho}}\Bigg)^{i+j} \frac{\pi}{n} \sum_{l=1}^n      \nonumber  \\
&~~~(\cos(\frac{2l-1}{2n}\pi)+1)^{i+j-1} \times e^{-\frac{2 \frac{a_w +a_y/a_2}{\rho}}{\cos(\frac{2l-1}{2n}\pi)+1}}  |\sin(\frac{2l-1}{2n}\pi)|.   
\end{align}

Thus, the approximated achievable rates (\ref{c_1_appro}) and (\ref{c_2_appro}) can be conveniently calculated numerically, and their accuracy will be validated by the simulation results in the next section.
\vspace*{-2mm}

\begin{figure}[htbp]
\centering
\includegraphics[width=3.3in,height=2.2in]{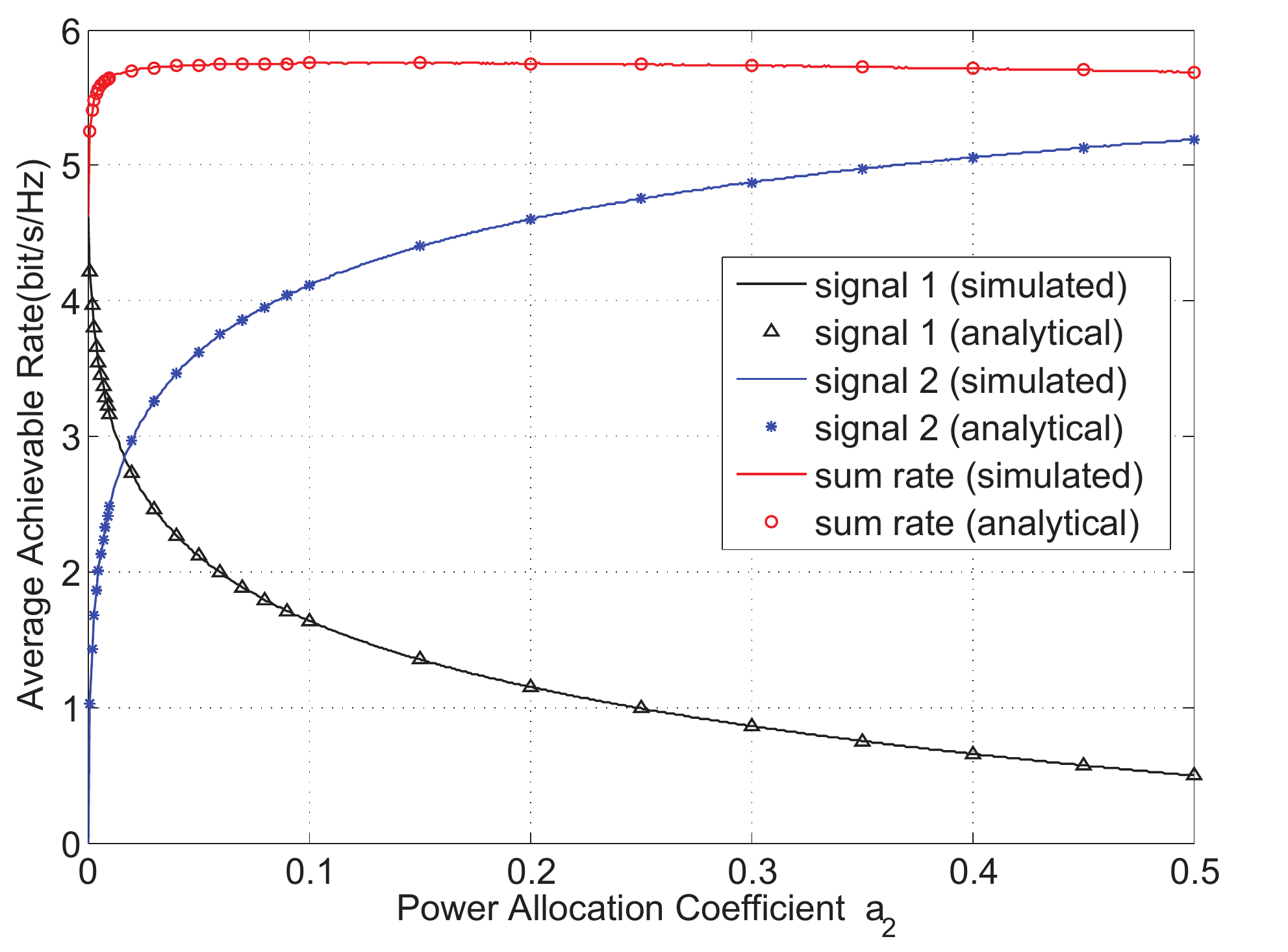}
\vspace*{-2mm}
\caption{Achievable rates for the NOMA-based CRS over Rician fading channels.}
\label{simulation_1}
\vspace*{-2mm}
\end{figure}
\vspace*{-4mm}
\begin{figure}[htbp]
\centering
\includegraphics[width=3.1in,height=2.2in]{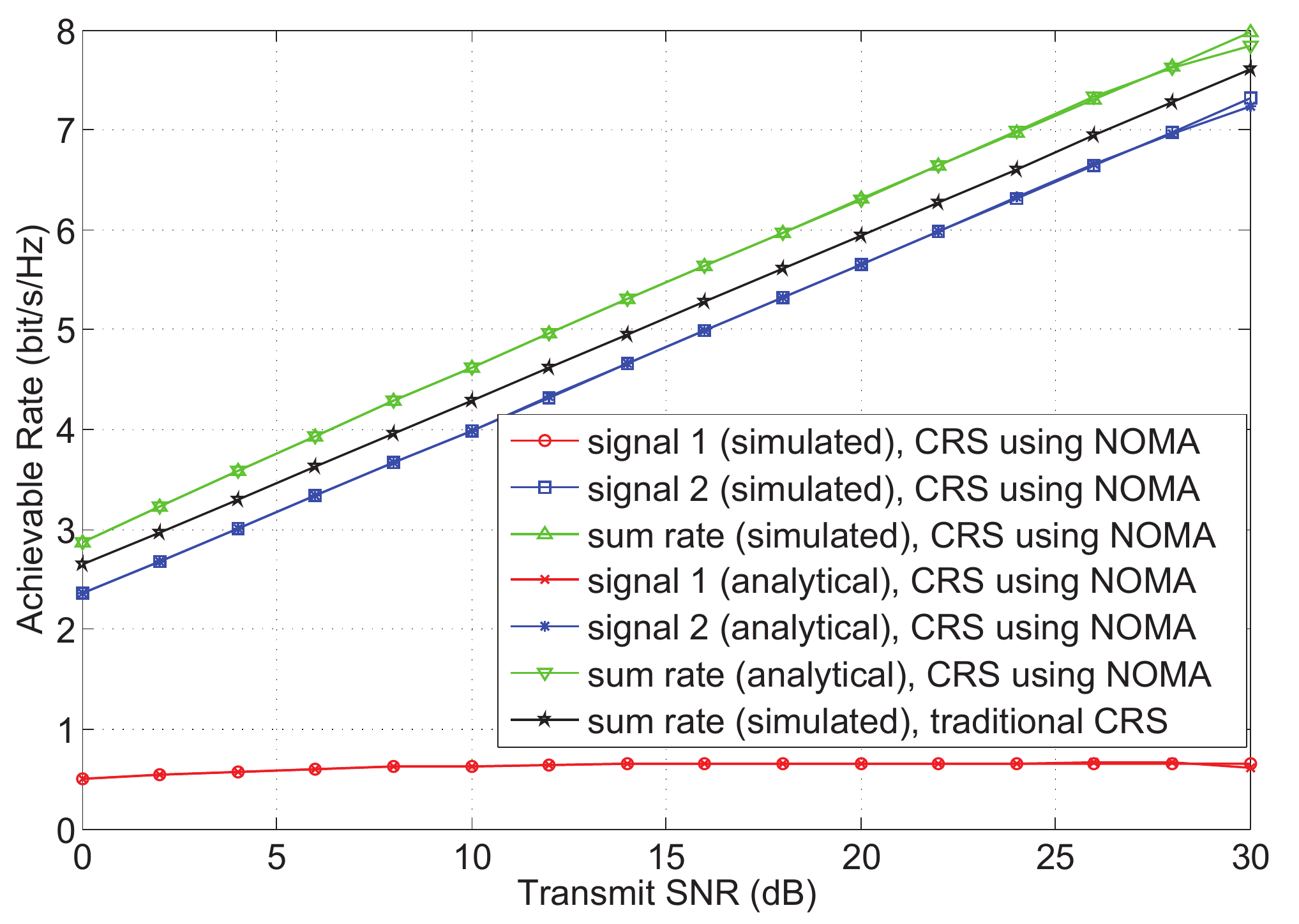}
\vspace*{-2mm}
\caption{Achievable rates comparison between the NOMA based CRS and traditional CRS.}
\label{simulation_2}
\vspace*{-2mm}
\end{figure}

\vspace*{2mm}

\section{Numerical Results and Simulations}

In this section, we compare the analytical results obtained in the previous Section III with Monte Carlo simulations to validate their accuracy. Specifically, $10^5$ realizations of Rician distribution random variables are generated, and the approximation order for Gauss-Chebyshev Integration is set as 100.

Fig. \ref{simulation_1} presents the achievable rate performance of $s_1$, $s_2$ and the corresponding sum rate of the NOMA-based CRS against the power allocation coefficient $a_2$. 
In the model of Rician fading channel, the parameter $\Omega$ denotes the average power gain of the channel \cite{rician}, which is usually determined by distances between transceivers. 
In our system, $\Omega_i$ ($i \in \{\rm{SD},\rm{SR},\rm{RD}\}$) denotes the average power gain of link \rm{SD}, link \rm{SR}, and link \rm{RD}, mainly reflecting the impacts of distances from S to D, from S to R, and from R to D, respectively. Thus, we set $\Omega_{\rm{SD}}=9<\Omega_{\rm{SR}}=\Omega_{\rm{R}D}=36$ \cite{kim2015capacity}, because the distances from S to R and from R 
to D are usually smaller than the distance from S to D, and thus link \rm{SR} and link \rm{RD} have higher average power gains than link \rm{SD}. According to \cite{bhatnagar2013capacity}, other parameters are set as SNR=20 dB, $K_{\rm{SR}}=K_{\rm{RD}}=5$, and $K_{\rm{SD}}=2$. From Fig. \ref{simulation_1}, we can observe that the derived analytical results using Gauss-Chebyshev Integration match well with the simulation results. In addition, as $a_2$ increases, $s_2$ will get more power and its achievable rate increases accordingly, while the achievable rate of $s_1$ decreases. Moreover, the sum rate of two signals first increases and then slowly decreases with the increase of $a_2$.  Actually, we can see from (10) that when $a_2$ is small, the achievable rate of $s_2$ is mainly determined by link \rm{SR}, as $a_2\lambda_{\rm{SR}}$ will always be smaller than $\lambda_{\rm{RD}}$. Due to SIC, $s_2$ will have no interference in link \rm{SR}, so increasing $a_2$ will largely increase the achievable rate of $s_2$, and thus increase the sum rate. However, when $a_2$ increases, $a_2\lambda_{\rm{SR}}$ will be larger than $\lambda_{\rm{RD}}$ at last, and the rate of link \rm{RD} will slowly become the determinant factor, which is not influenced by $a_2$. As a result, the increase of $s_2$'s achievable rate finally cannot make up for the decrease of $s_1$'s achievable rate, which causes the decrease of the sum rate, as shown in Fig. \ref{simulation_1}. Thus, there exists an optimal power allocation coefficient to maximize the sum rate, which is an interesting research topic deserving further investigation in the future.

Fig. \ref{simulation_2} compares the achievable rates of the traditional CRS and the NOMA-based CRS against the transmit SNR, where we set $a_2 = 0.4$, $\Omega_{SD}=9$, $\Omega_{RD}=36$, and $\Omega_{SR}=144$. We find that the simulation results and analytical results are consistent, and the NOMA-based CRS achieves higher achievable sum rate than the traditional CRS, since NOMA-based CRS can transmit two signals in two slots, while traditional CRS can only transmit one signal during the same time.

\vspace*{-2mm}

\section{Conclusions}

In this paper, we have investigated the performance of a NOMA-based cooperative relaying system by deriving the exact analytical expressions of the achievable rates. Moreover, an efficient approximation method using Gauss-Chebyshev Integration for the achievable rates was also proposed, which enables the sum series of the achievable rate expressions converge quickly. Simulation results have verified that our derived analytical results match well with the Monte Carlo simulations, and the NOMA-based CRS is able to achieve higher achievable rate than the traditional CRS.


%





\ifCLASSOPTIONcaptionsoff
  \newpage
\fi

\end{document}